\documentclass[11pt]{article}
\usepackage{amsmath}
\usepackage{amssymb}
\usepackage{amsfonts}
\usepackage{amscd}
\usepackage{delarray}
\usepackage{graphicx}

\def\bd{\begin{displaystyle}}
\def\ed{\end{displaystyle}}
\def\ba{\begin{eqnarray}}
\def\ea{\end{eqnarray}}
\def\bea{\begin{eqnarray*}}
\def\eea{\end{eqnarray*}}

\let\BBox\Box
\def\Box{$\BBox$}

\begin{document}

\title{A new formalism for the estimation of the $CP$-violation parameters}
\date{}
\author{}
\maketitle \vglue -1.8truecm \centerline{M. Courbage\footnote {\it
Laboratoire Mati\`ere et Syst\`emes Complexes (MSC),  UMR 7057 CNRS
et Universit\'e Paris 7- Denis Diderot, Case 7056, B\^{a}timent
Condorcet, 10, rue Alice Domon et Léonie Duquet 75205 Paris Cedex
13, FRANCE. \texttt{email: maurice.courbage@univ-paris-diderot.fr}},
T. Durt\footnote{TENA-TONA Vrije Universiteit Brussel, Pleinlaan 2,
B-1050 Brussels, Belgium. \texttt{email: thomdurt@vub.ac.be}} and
S.M. Saberi Fathi\footnote{Laboratoire de Physique Th\'eorique et
Mod\`elisation (LPTM), UMR 8089, CNRS et Universit\'e de
Cergy-Pontoise, Site Saint-Matrin, 2, rue Adolphe Chauvin, 95302
Cergy-Pontoise Cedex, FRANCE. \texttt{email:
majid.saberi@u-cergy.fr}} }

\vspace{1cm}

\begin{abstract}
In this paper, we use the time super-operator formalism in the
2-level Friedrichs model \cite{fried} to obtain a phenomenological
model of mesons decay. Our approach provides a fairly good
estimation of the $CP$ symmetry violation parameter in the case of
K, B  and D mesons. We also propose a crucial test aimed at
discriminating between the standard approach and the time
super-operator approach developed throughout the paper.
\end{abstract}

\bigskip




\section{Introduction}
There have been several theoretical approaches to $CP$ violation in
kaons (see e.g, the collection of papers edited in \cite{wolf}) and
the question is partially open today. In this paper, we use a
Hamiltonian model, describing a two-level states coupled to a
continuum of degrees of freedom, that makes is possible to simulate
the phenomenology of neutral kaons. Then, the time super-operator
formalism for the decay probability provides new numerical estimate
of the parameters of $CP$ violation.

It is well known \cite{perkins} that kaons appears in pair
$\mathrm{K}^0$ and $\overline{\mathrm{K}}^0$ each one being
conjugated to each other. The decay processes of $\mathrm{K}^0$ and
$\overline{\mathrm{K}}^0$  correspond to combinations of two
orthogonal decaying modes $\mathrm{K}_1$ and $\mathrm{K}_2$, that
are distinguished by their lifetime. The discovery of the small
$CP$-violation effect was also accompanied by the non orthonormality
of the short and long lived decay modes, now denoted $\mathrm{K}_S$
and $\mathrm{K}_L$, slightly different from $\mathrm{K}_1$ and
$\mathrm{K}_2$ and depending on a $CP$-violation parameter
$\epsilon$.  Lee, Oehme and Yang (LOY) \cite{loy} proposed a
generalization of the Wigner-Weisskopf theory \cite{ww} in order to
account the ``exponential decay". Later on, L. A. Khalfin
\cite{khalfin57} has pointed that, for a quantum system with energy
spectrum bounded from below, the decay could not be exponential for
large times. It was also observed \cite{girad} that short-time
behavior of decaying systems could not be exponential and this led
to the so-called Zeno effect \cite{chmisrasud,misrasud}. The
departure from the exponential type behavior has been experimentally
observed (see references quoted in \cite{chsud}).  L.A. Khalfin also
corrected the parameter $\epsilon$ at the lowest order of
perturbation. His estimation has been presented and reexamined in
the reference \cite{chsud} and applied to other mesons.

We show that our model allows us to obtain a better estimation of
the $CP$-violation parameter for kaon as well as  B and D mesons. We
also make new predictions that differ from standard predictions and
that could be tested experimentally.

This paper is organized as follows. In Section \ref{section2}, we
introduce the time super-operator for decay probability density.
Then, in Section \ref{friedrichssection} we present the 2-levels
Friedrichs model. Kaon phenomenology is recalled in Section
\ref{kaonpheno}. In Section \ref{WW}, we present the theory of
$CP$-violation in the Hilbert space and another derivation of the
intensity formula for mesons that already  has been used in
\cite{cds3}. Finally, in Section \ref{sec4},  we derive the time
super-operator intensity formula and we compute  $CP$-violation
parameters for K, B, and D mesons. Then, we compare our results with
the experimental data.

\section{Decay probability in the time super-operator ($T$)  approach}\label{section2}

\subsection{Decay probability in the time operator ($T'$)  approach
} In the Wigner-Weisskopf approximation to time evolution of quantum
unstable systems, the energy spectrum of the Hamiltonian is extended
from $-\infty$ to $+\infty$. In this approximation, a decay time
operator $T'$ is canonically conjugated to $H$. That is,
\begin{equation}\label{dc1}
H\psi(\omega)=\omega\psi(\omega)
\end{equation}
\begin{equation}\label{dc2}
T'\psi(\omega)=-\mathrm{i}\frac{d}{d\omega}\psi(\omega)
\end{equation}
so that $T'$ satisfies to the commutation relation
$[H,T']=\mathrm{i}I$. The $T'$-representation is obtained by a
Fourier transform
\begin{equation}\label{dc3}
\widehat{\psi}(\tau)=\frac{1}{\sqrt{2\pi}}\int_{-\infty}^\infty\,\,
e^{-\mathrm{i}\tau\omega}\, \psi(\omega)\, d\omega
\end{equation}
and the unstable states are the those prepared such that the decay
occurs in the future, that is, $\widehat{\psi}(\tau)=0$ for
$\tau<0$. Any state of the form
$\psi_{\mathrm{un}}(\omega)=A/(\omega-z_0)$, $(z_0=a-\mathrm{i}b,
b>0)$,  belongs to this space, since,
\begin{equation}\label{dc4}
\widehat{\psi}_{\mathrm{un}}(\tau)=\left\{\begin{array}{cc}
 \mathrm{i}A\sqrt{2\pi}e^{-\mathrm{i}\tau z_0} & \tau\geq0, \\
 0 & \tau<0
\end{array}\right.
\end{equation}
It is clear that these states correspond to a decay probability
density:
\begin{equation}\label{dc5}
    |\widehat{\psi}_{\mathrm{un}}(\tau)|^2=2\pi|A|^2 e^{-b\tau}
\end{equation}
This is an exponential distribution of decay times that is very
common in particle physics.


\subsection{ Time super-operator ($T$)
formalism}

Rigourously speaking, when the Hamiltonian has a positive spectrum,
it is forbidden in principle to define a time operator that
satisfies the commutation relation $[H, T'] = \mathrm{i}I$. This
argument was elaborated by Pauli who showed that if one could find
such an operator $\hat{T}'$ one could use it for generating
arbitrary translations in the energy eigenspace so that then the
spectrum of $\hat H$ ought to be unbounded by below, which clearly
constitutes a physical impossibility.

In order to escape this contradiction one needs to go to the space
of density matrices in order to obtain a time operator that is
conjugated to the evolution operator (the Liouville-von Neumann
operator) because it is sufficient that the Hamiltonian is not
upperly bounded so that the Liouville-von Neumann operator has a
spectrum extending from $-\infty$ to $\infty$. In order to do so,
let us consider the Liouville-von Neumann space which is the space
of operators $\rho $ on $ \mathcal{H}$ equipped with the scalar
product $<\rho, \rho'> = Tr(\rho^* \rho ') $ for which the time
evolution is given by
\begin{equation}
U_t\rho = e^{-itH}\rho e^{itH}
\end{equation}
  $U_t = e^{-itL}$ is generated by the Liouville von-Neumann operator
  $L$ given by:
\begin{equation}\label{L}
L\rho = H\rho - \rho H
\end{equation}
The time super-operator $T$ is a self-adjoint super-operator on the
Liouville-von Neumann space conjugated to $L$, i.e.
$[T,L]=\mathrm{i}I$. This definition is equivalent to the Weyl
relation: $e^{itL } T e^{-itL } = T + tI $.

The average of $T$ in the state $\rho$ is given by \footnote{ The
linearity that usually characterizes the relation between average
values of observable $A$ and density matrix $M$: $\mathrm{tr}(MA)$
seems to be violated here, but one should not forget that (a) in the
case of pure states the density matrix equals its square and (b)
this paradox is easily solved in the case of mixtures by imposing
that $\rho$ is the square root of the density matrix $M=\rho^*\rho$,
$\mathrm{tr}(MA)=\mathrm{tr}(\rho^*A\rho)$.}:
\begin{equation}
\langle  T \rangle _\rho = \langle \rho, T\rho\rangle
\end{equation}

 The time of occurrence of a random event fluctuates and we speak
of the probability of its occurrence in a time interval $I =]t_1,
t_2]$. The observable $T'=-T$ is associated to such event. In fact,
for a system in the initial state $\rho_0$ the average time of
occurrence $\langle T'\rangle_{\rho_0}$  is to be related to the
time parameter $t$ and to  the average time of occurrence in the
state $\rho_t=e^{-\mathrm{i}tL}\rho_0$   by:
\begin{equation}\label{tio18}
\langle T'\rangle_{\rho_t}=\langle T'\rangle_{\rho_0}-t
\end{equation}
This equation follows from the Weyl relation.

Let $ \mathcal{P}_\tau$ denote the family  of spectral projection
operators of $T$:
\begin{equation}
T = \int_{\mathbb{R}} \tau d\mathcal{P}_{\tau}
\end{equation}
and let $\mathcal{Q}_\tau$ be the family of spectral projections of
$T'$, then, in the state $\rho$, the probability of occurrence of
the event in a time interval $I$ is given, as in the usual
formulations, by
\begin{equation}\label{tio18-1}
\mathcal{P}(I,\rho)=\|\mathcal{Q}_{t_2}\rho\|^2-\|\mathcal{Q}_{t_1}\rho\|^2=
\|(\mathcal{Q}_{t_2}-\mathcal{Q}_{t_1})\rho\|^2:=\|\mathcal{Q}(I)\rho\|^2
\end{equation}
The unstable ``undecayed" states observed at $t_0=0$ are the states
$\rho$ such that $\mathcal{P}(I, \rho) = 0$ for any negative time
interval $I$, that is:
\begin{equation}\label{tio19}
\|\mathcal{Q}_{\tau}\rho\|^2=0, ~~\forall\tau\leq0
\end{equation}
In other words, these are the states verifying $\mathcal{Q}_0\rho =
0$. It is straightforwardly checked that the spectral projections
$\mathcal{Q}_{\tau}$ are related to the spectral projections
$\mathcal{P}_{\tau}$ by the following relation:
\begin{equation}\label{tio20}
\mathcal{Q}_{\tau} =1-\mathcal{P}_{-\tau}
\end{equation}
Let $\mathfrak{F}_{\tau}$ be the subspace on which
$\mathcal{P}_\tau$ projects . Thus, the unstable undecayed states
are those states satisfying $\rho = \mathcal{P}_0\rho$ and they
coincide with the subspace $\mathfrak{F}_0$\footnote{Therefore, a
subspace $\mathfrak{F}_{t_0}$ is a set of decaying states prepared
at time $t_0$. We call it an  unstable space of T.}. For these
states, the probability that a system prepared in the undecayed
state $\rho$ is found to decay some time during the interval $I =]0,
t]$ is $\|\mathcal{Q}_{t}\rho\|^2=1-\|\mathcal{P}_{-t}\rho\|^2$ a
monotonically nondecreasing quantity which converges to $1$ as
$t\rightarrow\infty$ while $\|\mathcal{P}_{-t}\rho\|^2$ tends
monotonically to zero. As noticed by Misra and Sudarshan
\cite{misrasud}, such quantity can not exist in the usual quantum
mechanical treatment of the decay processes. It should not be
confused with the usual ``survival probability of an unstable state
$\chi$ at time $t$ " defined by $\mid<\chi, e^{-itH}\chi>\mid ^2$
where $\chi$ is  an eigenstate of the free Hamiltonian. In fact,
{\it the last quantity is interpreted as the probability,  at the
instant t}, for finding the system undecayed when at time 0 it was
prepared in the state $\chi$. There is no general reason for this
quantity to be  monotonically decreasing  as should be any genuine
probability distribution.  This problem does not appear in the time
operator approach.

Considered so, the time operator approach is non-standard. Actually,
the key, non-standard, assumption that underlies the time
super-operator formalism is that.

 {\it In the Liouville space, given
any initial state $\rho$, its survival probability in the unstable
space is given by:}
\begin{equation}\label{tio21}
p_\rho(t)=\|\mathcal{P}_{0}e^{-\mathrm{i}tL}\rho\|^2
\end{equation}
\emph{This is the probability that, for a system initially in the
state $\rho$, no decay is found during} $[0,t]$. Given any initial
state $\rho$, its survival probability in the unstable space is
given by \cite{courbage}
\begin{eqnarray}\label{tio22}
\nonumber p_{\rho}(t) &=& \| \mathcal{P}_0e^{-\mathrm{i}tL}\rho \|^2\\
\nonumber &=&\| U_{-t}\mathcal{P}_0U_t\rho \|^2\\
&=& \| \mathcal{P}_{-t}\rho \|^2
\end{eqnarray}
Here we used the following relation:
$\mathcal{P}_{-t}=U_{-t}\mathcal{P}_0 U_t$. \emph{Then, the survival
probability  is monotonically decreasing to $0$ as }$t \rightarrow
\infty$.  This survival probability and the probability of finding
the system to decay some time during the interval $I = ]0, t],~
q_\rho(t)=\|\mathcal{Q}_\rho(t)\|^2$ are related by:
\begin{equation}\label{d-s}
q_{\rho}(t)= 1 - p_{\rho}(t)
\end{equation}
Therefore, $q_\rho(t)\rightarrow1$ when $t\rightarrow+\infty$.

\bigskip
The expression of the time operator is given in a {\it spectral
representation of} $H$, that is, in the representation in which $H$
is diagonal. As shown in \cite{cour80}, $H$ should have an unbounded
absolutely continuous spectrum. In the simplest case, we shall
suppose that $H$ is represented as the multiplication operator on
${\cal H}=L^{2}(\mathbb{R}^+)$ :
\begin{equation}
H\psi(\omega)=\omega\psi(\omega).
\end{equation}
The Hilbert-Schmidt operators on $L^{2}(\mathbb{R}^+)$ correspond to
the square-integrable functions $ \rho(\omega,\omega^{'})\in
L^{2}(\mathbb{R}^+ \times \mathbb{R}^+ )$ and the Liouville-von
Neumann operator $L$ is given by :
\begin{equation}\label{l1}
L \rho(\omega,\omega^{'})=
(\omega-\omega^{'})\rho(\omega,\omega^{'})
\end{equation}
Then we obtain a spectral representation of $L$ via the change of
variables:
\begin{equation}\label{l2}
\nu = \omega-\omega^{'}
\end{equation}
and
\begin{equation}\label{l3}
E= \min(\omega,\omega^{'})
\end{equation}
This gives a spectral representation of
 $L$:
\begin{equation}
L\rho(\nu, E)= \nu \rho(\nu, E),
\end{equation}
where $ \rho(\nu,E)\in L^{2}(\mathbb{R} \times \mathbb{R}^+ )$. In
this representation, $T \rho(\nu, E)= \mathrm{i}
\frac{d}{d\nu}\rho(\nu, E)$ so that the spectral representation, of
$T$ is obtained by the inverse Fourier transform:
\begin{equation}\label{inf}
\hat{\rho}(\tau, E) = \frac{1}{\sqrt{2 \pi}}
\int_{-\infty}^{+\infty} e^{\mathrm{i}\tau \nu}\rho(\nu, E)d\nu =(
{\cal F}^* \rho)(\tau, E)
\end{equation}
and
\begin{equation}
T\hat{\rho}(\tau, E) = \tau \hat{\rho}(\tau, E).
\end{equation}
The spectral projection operators ${\cal P}_s$ of $T$ are given in
the $(\tau, E)$-representation by
\begin{equation}
{\cal P}_s\hat{\rho}(\tau, E)=
\chi_{]-\infty,s]}(\tau)\hat{\rho}(\tau, E)
\end{equation}
where $\chi_{ ]-\infty,s]}$ is the characteristic function of
$]-\infty,s]$.  So, to obtain in the $(\nu,E)$-representation the
 expression of these spectral projection operators, we use the
 Fourier transform:
\begin{eqnarray}
\nonumber{\cal P}_s\rho(\nu, E)&=&
\frac{1}{\sqrt{2\pi}}\int_{-\infty}^{s}  e^{-\mathrm{i} \nu
\tau}\hat\rho(\tau, E)\,d\tau\\&=&e^{-\mathrm{i}\nu
s}\int_{-\infty}^{0} e^{-i\nu\tau}\hat\rho(\tau+s, E)\,d\tau.
\label{proj1}
\end{eqnarray}
Let $ g\in L^2(\mathbb{R}) $ and  denote its Fourier transform by:
${\cal F}g(\nu)=\frac{1}{\sqrt{2\pi}}\int_{-\infty}^{\infty}
e^{-\mathrm{i} \nu \tau}g(\tau)\,d\tau $.  Using the Hilbert
transformation:
\begin{equation}
\mathbf{H}g(x)=\frac{1}{\pi}\textsf{P}\int_{-\infty}^{\infty}\frac{g(t)}{t-x}\,dt.
\end{equation}
We have \cite{titch} the following formula:
\begin{equation}
\frac{1}{\sqrt{2\pi}}\int_{-\infty}^{0}  e^{-\mathrm{i} \nu
\tau}g(\tau)\,d\tau=\frac{1}{2}({\cal F}(g)-\mathrm{i}
\mathbf{H}{\cal F}(g)).\label{proj2}
\end{equation}
Finally, using the well-known property of the translated Fourier
transform: $\sigma_s g(\tau)=g(\tau+s)$,
\begin{equation}
{\cal F}(\sigma_s g)(\nu)=e^{\mathrm{i} \nu s}{\cal F}.g(\nu)
\end{equation}
(\ref{proj1}) and (\ref{proj2}) yield:
\begin{equation}
{\cal P}_s\rho(\nu,E)=\frac{1}{2}e^{-\mathrm{i}\nu s}[e^{i\nu
s}\rho(\nu,E)-\mathrm{i} \mathbf{H}(e^{\mathrm{i}\nu
s}\rho(\nu,E))].
\end{equation}
Thus:
\begin{equation}
{\cal P}_s\rho(\nu,E)=\frac{1}{2}[\rho(\nu,E)-\mathrm{i}
e^{-\mathrm{i}\nu s}\mathbf{H}(e^{\mathrm{i}\nu
s}\rho(\nu,E))].\label{t1}
\end{equation}
It is to be noted that ${\cal P}_s\rho(\nu,E)$ is in the Hardy class
$ \mathbb{H}^+$ (i.e. it is the limit as $y \rightarrow0^+ $ of an
analytic function $\Phi(\nu+\mathrm{i} y)$ such that:
$\int_{-\infty}^{\infty}\mid\Phi(\nu+\mathrm{i}\emph{y})\mid^2\,dy<\infty$)\cite{titch}.


\section{The two-level Friedrichs model}\label{friedrichssection}
The Friedrichs interaction Hamiltonian between the two discrete
modes and the continuous degree of freedom is given by the operator
$H$ on the Hilbert space of the wave functions of the form
$\mid\psi>=\{f_1,f_2,g(\mu)\}, f_1,f_2\in\mathbb{C}, g\in
L^2(\mathbb{R}^+)$
\begin{equation}
H=H_0 + \lambda_1 V_1+\lambda_2 V_2,
\end{equation}
where $\lambda_1$ and $\lambda_2$ are the complex coupling
constants, and
\begin{equation}
H_0 \mid \psi>=\{\omega_1 f_1,\omega_2 f_2,\mu g(\mu)\}, (\omega_1
~~\mathrm{and}~~ \omega_2 > 0).
\end{equation}

The operators $V_i ~(i=1,2)$ are given by:
\begin{eqnarray}
\nonumber V_1\{f_1,f_2,g(\mu)\}=\{<v(\mu),g(\mu)>,0,f_1.v(\mu)\}\\
V_2\{f_1,f_2,g(\mu)\}=\{0,<v(\mu),g(\mu)>,f_2.v(\mu)\}
\end{eqnarray}
where
\begin{equation}
<v(\mu),g(\mu)>=\int d\mu v^*(\mu)g(\mu),
\end{equation}
is the inner product. Thus $H$ can be represented as a matrix :

\begin{equation}
H_{\mathrm{Friedrichs}}=\left(\begin{array}{ccc} \omega_1 & 0&
\lambda_1^* v^*(\mu)  \\ 0& \omega_2 & \lambda_2^* v^*(\mu)
\\ \lambda_1
v(\mu)& \lambda_2 v(\mu)& \mu
\end{array}\right)\label{hamil}
\end{equation}
$\omega_{1,2}$ represent the energies of the discrete levels, and
the factors $\lambda_{i} v(\mu)~ (i=1,2)$ represent the couplings to
the continuous degree of freedom. The energies $\mu$ of the
different modes of the continuum range from $-\infty$ to $+\infty$
when $v(\mu)=1$, but we are free to tune the coupling $v(\mu)$ in
order to introduce a selective cut off to extreme energy modes. Let
us now solve the Schr\"{o}dinger equation and trace out the
continuum in order to derive the master equation for the two-level
system. The two-level Friedrichs model Schr\"{o}dinger equation with
$\hbar=1$ is formally written as
\begin{equation}
\left(\begin{array}{ccc} \omega_1 & 0& \lambda_1^* v^*(\mu)  \\ 0&
\omega_2 & \lambda_2^* v^*(\mu)\\ \lambda_1 v(\mu)& \lambda_2
v(\mu)& \mu
\end{array}\right)\left(\begin{array}{ccc}f_1 \\ f_2 \\ g(\mu)
\end{array}\right)=\omega\left(
\begin{array}{ccc}f_1 \\ f_2 \\ g(\mu)
\end{array}\right).\label{f1}
\end{equation}
That is to say:
\begin{equation}
\omega_1 f_1(\omega)+\lambda_1^*\int d\mu v^*(\mu) g(\mu)=\omega
f_1(\omega) ,~\label{af2}
\end{equation}
\begin{equation}
\omega_2 f_2(\omega)+\lambda_2^*\int d\mu v^*(\mu) g(\mu)=\omega
f_2(\omega),\label{f3}
\end{equation}
and
\begin{equation}
\lambda_1 v(\omega)f_1(\omega)+\lambda_2 v(\omega)f_2(\omega)+\mu
g(\omega)=\omega g(\omega).\label{f4}
\end{equation}
The solution of (\ref{f4}), for ``outgoing" wave, is:
\begin{eqnarray}
g(\mu)=\delta(\mu-\omega)-\lim_{\epsilon\rightarrow
0}\frac{\lambda_1 v(\mu)f_1+\lambda_2
v(\mu)f_2}{\omega-\mu-\mathrm{i}\epsilon}.
\end{eqnarray}
inserting the above equation in the equations(\ref{af2}) yields
\begin{eqnarray}
f_1(\omega)=\frac{\lambda_1^*
v^*(\omega)}{\eta_1^+(\omega)}-\bigg{(}\lambda_1^*\lambda_2\lim_{\epsilon\rightarrow
0}\int
d\mu\frac{|v(\mu)|^2}{\mu-\omega-\mathrm{i}\epsilon}\bigg{)}f_2(\omega),~\label{af3}
\end{eqnarray}
where
\begin{eqnarray}
\eta_1^+(\omega)=\omega-\omega_1+|\lambda_1|^2\lim_{\epsilon\rightarrow
0}\int d\mu\frac{|v(\mu)|^2}{\mu-(\omega+\mathrm{i}\epsilon)}.
\end{eqnarray}
We can also obtain the similar relations for $f_2$ by changing the
indices $1$ with $2$ and vis versa as:
\begin{eqnarray}\label{af3-1}
f_2(\omega)=\frac{\lambda_2^*
v^*(\omega)}{\eta_2^+(\omega)}-\bigg{(}\lambda_1\lambda_2^*\lim_{\epsilon\rightarrow
0}\int
d\mu\frac{|v(\mu)|^2}{\mu-\omega-\mathrm{i}\epsilon}\bigg{)}f_1(\omega).
\end{eqnarray}
By substituting $f_2(\omega)$ from the above equation in the
equation (\ref{af3}) we obtain
\begin{eqnarray}\label{af3-2}
\nonumber f_1(\omega)&=&\frac{1}{1-\left(\lambda_1^*\lambda_2\int
d\mu\frac{|v(\mu)|^2}{\mu-\omega-\mathrm{i}0}\right)^2}\left(\frac{\lambda_1^*
v^*(\omega)}{\eta_1^+(\omega)}-\frac{\lambda_1^*|\lambda_2|^2}{\eta_2^+(\omega)}\int
d\mu\frac{|v(\mu)|^2}{\mu-\omega-\mathrm{i}0}\right)\\
&=&\frac{1}{1-O(|\lambda|^4)}\bigg{(}\frac{\lambda_1^*
v^*(\omega)}{\eta_1^+(\omega)}-O(\lambda_1^*|\lambda_2|^2)\bigg{)}
\end{eqnarray}
Thus, to the order two approximation we have
\begin{equation}\label{af4}
f_1(\omega)\simeq\frac{\lambda_1^* v^*(\omega)}{\eta_1^+(\omega)}.
\end{equation}
and the same formula for $f_2$ as:
\begin{equation}\label{af4-1}
f_2(\omega)\simeq\frac{\lambda_2^* v^*(\omega)}{\eta_2^+(\omega)}.
\end{equation}
 Also denote
$\eta_i^-(\omega)=\eta_i(\omega -\mathrm{i} \epsilon)$.
$\eta_i^\pm(\omega),~(i=1,2)$ are complex conjugate of each other,
we can see that
\begin{equation}
\eta_i^\pm(\omega)=\omega - \omega_i + |\lambda_i|^2~
\textsf{P}\int_0^{\infty}
\frac{|v(\omega^{'})|^2}{\omega^{'}-\omega}~d\omega^{'}\pm\mathrm{i}\pi
 |\lambda_i|^2|v(\omega)|^2,\label{pole}
\end{equation}
where $\textsf{P}$ indicates the ``principal value" and we used the
following identity in equation (\ref{pole})
\begin{equation}
\lim_{\varepsilon\rightarrow
0^+}\frac{1}{x-x_0\pm\mathrm{i}\varepsilon}=\textsf{P}\frac{1}{x-x_0}\mp\mathrm{i}\pi\delta(x-x_0).
\end{equation}

Let $|\chi\rangle=|\epsilon_1f_1+\epsilon_2 f_2\rangle$ where
$\epsilon_i,~~~(i=1,2)$ is a constant complex number.  The physical
meaning of such a state is that it corresponds to a coherent
superposition of two exponential decay processes. In the following
Section we shall compute the projection of
$|\chi\rangle\langle\chi|$ on the unstable spaces of time operator
and then the survival probability $p_{\rho}(t)$ introduced in the
Section 2. We compute its expression for the density matrix
$|\chi\rangle\langle\chi|$ in terms of the lifetimes and energies of
the (mesonic) resonances. It has been shown \cite{ordonez} that the
average of time operator for the state $|\chi\rangle\langle\chi|$ is
equal to the lifetime $ 1/\gamma$ in a first approximation (more
precisely in the weak coupling regime that is described in the next
section (equation (\ref{approxpole})). We shall characterize the
short time and long time behavior of this survival probability.

 Let us now identify the pure state $\chi$ with
the element $\rho= |\chi><\chi|$ of the Liouville space, that is the
kernel operator:
\begin{equation}
\rho=\sum_{i=1}^{2}\sum_{j=1}^2\rho_{ij}(\omega,
\omega^{'})=\sum_{i=1}^{2}\sum_{j=1}^2\epsilon_i\epsilon^*_jf_i(\omega)\overline{
f_j(\omega^{'}})=\sum_{i=1}^{2}\sum_{j=1}^2\epsilon_i\epsilon^*_j\mathfrak{F}_{ij}.
\end{equation}
We shall compute the survival probability $\|{\cal P}_{-s} \rho\|^2$
of the state $\rho$ and show how it reaches the following limit:
\begin{equation}
\lim_{s\rightarrow \infty}\|{\cal P}_{-s} \rho \|^2 \rightarrow 0  .
\end{equation}


\subsection{Weak coupling conditions}

 As explained above the Liouville operator is
given by equation (\ref{l1}) and the spectral representation of $L$
is given by the change of variables introduced in (\ref{l2}) and
(\ref{l3}). Thus, we obtain for $\mathfrak{F}_{ij}(\nu,E),
~~(i,j=1,2)$ :
\begin{equation}
\mathfrak{F}_{ij}(\nu,E)=\left\{ \begin{array}{ll}
\lambda_i\lambda_j^*\frac{v(E)}{\eta_i^{-}(E)}\frac{v^*(E+\nu)}{\eta_j^{+}(E+\nu)}&
\mbox {$\nu >0$}\\\\
\lambda_i^*\lambda_j\frac{v^*(E)}{\eta_j^{+}(E)}\frac{v(E-\nu)}{\eta_i^{-}(E-\nu)}&
\mbox{$\nu < 0$},\end{array} \right.
\end{equation}

Admitting that  $\eta_i^+(\omega)$ in (\ref{pole}) in the  the
$O(|\lambda|^2)$ has one zero in the lower half-plane
\cite{marchand,cs4} which approaches $\omega_i$ for decreasing
coupling, we can write:
\begin{eqnarray}\label{approxpole}
\eta_i^+(\omega)=  \omega -z_i.
\end{eqnarray}
where $z_i=\widetilde{\omega}_i-\mathrm{i}\frac{\gamma_i}{2}$ where
$\gamma_i\sim|\lambda_i|^2$ is a real positive constant \cite{cs4}.
In this article we suppose that
$\widetilde{\omega}_1<\widetilde{\omega}_2$. Easily, we can verify
that
\begin{equation}
\eta_i^+(\omega)-\eta_i^-(\omega)=\mathrm{i} \gamma_i.
\end{equation}
From (\ref{pole}), we have
\begin{eqnarray}\label{etap1}
\frac{\mathrm{i}}{2}
\left[\frac{1}{\eta_i^+(\omega)}-\frac{1}{\eta_i^-(\omega)}\right]=\frac{\pi|\lambda_i|^2
|v(\omega)|^2}{|\eta_i^+(\omega)|^2}.
\end{eqnarray}
Consequently, the two above equations yield
\begin{eqnarray}\label{etap2}
\frac{\pi|\lambda_i|^2 |v(\omega)|^2}{|\eta_i^+(\omega)|^2}=\frac{
\frac{\gamma_i}{2}}{|\eta_i^+(\omega)|^2}.
\end{eqnarray}
Therefore,
$|f_i(\omega)|^2\sim\frac{1}{(\omega-\widetilde{\omega}_i)^2+\frac{\gamma_i^2}{4}}$
which is a Breit-Wigner like distribution. This equation will be
used in the next sections.

\section{Phenomenology of kaons}\label{kaonpheno}
Kaons are bosons that were discovered in the forties during the
study of cosmic rays. They are produced by collision processes in
nuclear reactions during which the strong interactions dominate.
They appear in pairs $\mathrm{K}^{0}$, $\overline{\mathrm{K}}^{0}$
\cite{perkins,hokim}.

The $\mathrm{K}$ mesons are eigenstates of the parity operator $P$:
$ P|\mathrm{K}^0\rangle=- |\mathrm{K}^0\rangle$, and $
P|\overline{\mathrm{K}}^0\rangle=- |\overline{\mathrm{K}}^0\rangle$.
$\mathrm{K}^0$ and $\overline{\mathrm{K}}^0$ are charge conjugate to
each other  $ C|\mathrm{K}^0\rangle=
|\overline{\mathrm{K}}^0\rangle$, and $
C|\overline{\mathrm{K}}^0\rangle= |\mathrm{K}^0\rangle$. We get thus
\begin{equation}
CP|\mathrm{K}^0\rangle= -|\overline{\mathrm{K}}^0\rangle, \indent
 CP|\overline{\mathrm{K}}^0\rangle=
-|\mathrm{K}^0\rangle.
\end{equation}
Clearly $|\mathrm{K}^0\rangle$ and $|\overline{\mathrm{K}}^0\rangle$
are not $CP$-eigenstates, but the  following combinations
\begin{equation}
|\mathrm{K}_1\rangle=\frac{1}{\sqrt{2}}\big{(}|\mathrm{K}^0\rangle
+|\overline{\mathrm{K}}^0\rangle\big{)} , \indent
|\mathrm{K}_2\rangle=\frac{1}{\sqrt{2}}\big{(}|\mathrm{K}^0\rangle
-|\overline{\mathrm{K}}^0\rangle\big{)}\label{kk1},
\end{equation}
are $CP$-eigenstates.
\begin{equation}
CP|\mathrm{K}_1\rangle=+|\mathrm{K}_1\rangle, \indent
CP|\mathrm{K}_2\rangle=-|\mathrm{K}_2\rangle.
\end{equation}
In the absence of matter, kaons disintegrate through weak
interactions \cite{hokim}. Actually, $\mathrm{K}^0$ and
$\overline{\mathrm{K}}^0$ are distinguished by their mode of
\emph{production}. $\mathrm{K}_1$ and $\mathrm{K}_2$ are the decay
modes of kaons. In absence of $CP$-violation, the weak
disintegration process distinguishes the $\mathrm{K}_{1}$ states
which decay only into ``$2\pi$" while the $\mathrm{K}_{2}$ states
decay into ``$3\pi, \pi e \nu, ...$" \cite{leebook}. The lifetime of
the $\mathrm{K}_{1}$ kaon is short ($\tau_{S}\approx
8.92\times10^{-11}~^\mathrm{s}$), while the lifetime of the
$\mathrm{K}_{2}$ kaon is quite longer ($\tau_{L}\approx
5.17\times10^{-8}~^\mathrm{s}$).

\bigskip

$CP$-\emph{violation}  was discovered by Christenson \emph{et al}.
\cite{christ}. $CP$-violation means that the long-lived kaon can
also decay to ``$2\pi"$. Then, the $CP$ symmetry is slightly
violated (by a factor of the order of $10^{-3}$) by weak
interactions so that the $CP$ eigenstates $\mathrm{K}_1$ and
$\mathrm{K}_2$ are not exact eigenstates of the decay interaction.
Those exact states are characterized by lifetimes that are in a
ratio of the order of $10^{-3}$, so that they are called the
short-lived state ($\mathrm{K}_S$) and long-lived state
($\mathrm{K}_L$ ). They can be expressed as coherent superpositions
of the $\mathrm{K}_1$ and $\mathrm{K}_2$ eigenstates through
\begin{equation}\label{kk2}
|\mathrm{K}_L\rangle=\frac{1}{\sqrt{1+|\epsilon|^2}}\big{[} \epsilon
~|\mathrm{K}_1\rangle + |\mathrm{K}_2\rangle \big{]},~~
|\mathrm{K}_S\rangle=\frac{1}{\sqrt{1+|\epsilon|^2}}\big{[}
|\mathrm{K}_1\rangle +\epsilon ~ |\mathrm{K}_2\rangle \big{]},
\end{equation}
where $\epsilon$ is a complex $CP$-violation parameter,
$|\epsilon|\ll1$ and $\epsilon$ does not have to be real.
$\mathrm{K}_L$ and $\mathrm{K}_S$ are the eigenstates of the
Hamiltonian for the mass-decay matrix \cite{hokim,leebook} which has
 the following form in the basis $|\mathrm{K}^0\rangle$ and
$|\overline{\mathrm{K}}^0\rangle$:
\begin{equation}
H=M-\frac{\mathrm{i}}{2}\Gamma\equiv \left(\begin{array}{cc}
M_{11}-\frac{\mathrm{i}}{2}\Gamma_{11} &
M_{12}-\frac{\mathrm{i}}{2}\Gamma_{12}  \\
M_{21}-\frac{\mathrm{i}}{2}\Gamma_{21} &
M_{22}-\frac{\mathrm{i}}{2}\Gamma_{22}
\end{array}\right)
\label{massdecay}\end{equation} where $M$ and $\Gamma$ are
individually hermitian since they correspond to observables (mass
and lifetime). The corresponding eigenvalues of the mass-decay
matrix are equal to
\begin{equation}\label{kkeigen}
m_L-\frac{\mathrm{i}}{2}\Gamma_L, \indent
m_S-\frac{\mathrm{i}}{2}\Gamma_S
\end{equation}
The $CP$-violation was established by the observation that
$\mathrm{K}_L$ decays not only via three-pion, which has natural
$CP$ parity, but also via the two-pion (``$2\pi"$) mode with an
experimentally observed violation amplitude $|\epsilon^{exp.}|$ of
the order of $10^{-3}$, which was truly unexpected at the time. Let
us now reconsider how the simple model (\ref{kk2}),
(\ref{massdecay}) is related to the experimental data. A series of
detections is performed at various distances from the source of a
neutral kaon beam in order to estimate the variation of the
populations of emitted pion $\pi^+,\pi^-$ pairs in function of the
proper time. This is done for times of the order of $\tau_S$. The
experiment shows that an interference term is present in the
expression of the excitation rates of detectors in function of their
distance to the source.  It follows from (\ref{kk2}) that the
transition amplitude of the $\mathrm{K}_L$ beam is given by
\begin{equation}\label{psi}
\psi(t)=A\left( e^{-\mathrm{i} (m_S-\frac{\mathrm{i}}{2}\Gamma_S)
t}+\epsilon^{\mathrm{exp}} e^{-\mathrm{i}
(m_L-\frac{\mathrm{i}}{2}\Gamma_L) t}\right)
\end{equation}
with $A$ a global proportionality factor that remains constant in
time. Then the intensity $I(t)=|\psi(t)|^2$ is given by:
\begin{equation}\label{intens}
I(t)=I_0\,\left(e^{-\Gamma_S t}+ |\epsilon^{\mathrm{exp}}
|^2e^{-\Gamma_L t}+ |\epsilon^{\mathrm{exp}} |e^{-({\Gamma_S
+\Gamma_L\over 2})t)} \cos (\triangle m
t+\arg(\epsilon^{\mathrm{exp}}))\right)
\end{equation}
where
\begin{equation}\label{ratio}
|\epsilon^{\mathrm{exp}}| ={\mathrm{Amplitude }\,(K_L\rightarrow
\pi^+,\pi^-)
 \over \mathrm{Amplitude}\,(K_S\rightarrow \pi^+,\pi^-)}
 \end{equation}
 By fitting the expressions (\ref{intens}) and (\ref{ratio}) with the observed
data one derives an estimation of the mass difference between the
short and long lived state as well as the phase of
$\epsilon^{\mathrm{exp}}$ and its amplitude.

All this leads to an experimental estimation of
$\epsilon^{\mathrm{exp}}$ \cite{kk}
\begin{equation}\label{expepsilon}
|\epsilon^{\mathrm{exp}}|=(2.232\pm0.007)\times10^{-3}, ~~~
\mathrm{arg}(\epsilon^{\mathrm{exp}})=(43.5+0.7)^\circ.
\end{equation}

\section{The Wigner-Weisskopf type theory of the $CP$-violation in
the Hilbert space}\label{WW}

Let us present the fundamental ideas of the theory of spontaneous
emission  of an atom interacting with the electromagnetic field,
given by  Wigner and Weisskopf . This treatment aims at obtaining an
exponential time dependence for decaying states by integrating over
the continuum energy. That is, we assume that the modes of the
fields are closely spaced. Then, we have to assume that the
variation of  $v(\mu)$ over $\mu$ is negligible with $|\mu|\lesssim$
``uncertainty of the original state energy", i.e. $v(\mu)\approx v $
independent of $\mu$ or in the simple case it is taken to obey
$v(\mu)=1$. Also another assumption is that the lower limit of
integration over $\omega$ is replaced by $-\infty$.

 The two-level Friedrichs model
time-dependent Schr\"{o}dinger equation, in the Wigner-Weisskopf
regime becomes:
\begin{equation}
\left(\begin{array}{ccc} \omega_1 & 0& \lambda_1^*   \\ 0& \omega_2
& \lambda_2^* \\ \lambda_1 & \lambda_2 & \mu
\end{array}\right)\left(\begin{array}{ccc}f_1(t) \\ f_2(t) \\ g(\mu,t)
\end{array}\right)=\mathrm{i}\frac{\partial}{\partial t}\left(
\begin{array}{ccc}f_1(t) \\ f_2(t) \\ g(\mu,t)
\end{array}\right).\label{f1}
\end{equation}
which means:
\begin{equation}
\omega_1 f_1(t)+\lambda_1^*\int_{-\infty}^{\infty} d\mu
g(\mu,t)=\mathrm{i}\frac{\partial f_1(t)}{\partial t} ,~\label{f2}
\end{equation}
\begin{equation}
\omega_2 f_2(t)+\lambda_2^*\int_{-\infty}^{\infty} d\mu
g(\mu,t)=\mathrm{i}\frac{\partial f_2(t)}{\partial t},\label{f3}
\end{equation}
and
\begin{equation}
\lambda_1 f_1(t)+\lambda_2 f_2(t)+\mu
g(\mu,t)=\mathrm{i}\frac{\partial g(\mu,t)}{\partial t}.\label{f4}
\end{equation}
 Let us now solve the Schr\"odinger
equation and trace out the continuum in order to derive the master
equation for the two-level system. From the equation (\ref{f4})  we
can obtain $g(\mu,t)$, taking $g(\mu,0)=0$, as
\begin{equation}
g(\mu,t)=-\mathrm{i} e^{-\mathrm{i}\omega t}\int_0^t d\tau \big{[}
\lambda_1 f_1(\tau)+\lambda_2 f_2(\tau)\big{]}
 e^{\mathrm{i}\omega\tau},
\end{equation}
where $t>0$. Then, we substitute $g(\mu,t)$ in the equation
(\ref{f2}) and we obtain
\begin{equation}
\mathrm{i}\frac{\partial f_1(t)}{\partial t}=\omega_1
f_1(t)-\mathrm{i}\lambda_1^*\int_{-\infty}^{\infty} d\mu
e^{-\mathrm{i}\mu t}\int_0^t d\tau \big{[} \lambda_1
f_1(\tau)+\lambda_2 f_2(\tau)\big{]}
e^{\mathrm{i}\mu\tau},\label{f5}
\end{equation}
we also obtain the same relation for $f_2(t)$ from
equation(\ref{f3}):
\begin{equation}
 \mathrm{i}\frac{\partial f_2(t)}{\partial
t}=\omega_2 f_2(t)-\mathrm{i}\lambda_2^*\int_{-\infty}^{\infty} d\mu
e^{-\mathrm{i}\mu t}\int_0^t d\tau \big{[} \lambda_1
f_1(\tau)+\lambda_2 f_2(\tau)\big{]}
e^{\mathrm{i}\mu\tau}.\label{f6}
\end{equation}
 Finally, one obtains the following Markovian
form of the reduced Schr\"{o}dinger equation \cite{cds1}
\begin{equation}
\mathrm{i}\frac{\partial }{\partial t} \left(\begin{array}{c}
f_1(t)\\f_2(t)
\end{array}\right)=\left(
\begin{array}{cc}
\omega_1 - \mathrm{i} \pi |\lambda_1|^2 & - \mathrm{i} \pi
\lambda_1^*\lambda_2 \\
- \mathrm{i} \pi  \lambda_1\lambda_2^*& \omega_2 - \mathrm{i} \pi
|\lambda_2|^2
\end{array}\right)\left(
\begin{array}{c}
f_1(t)\\f_2(t)
\end{array}\right).\label{fh1}
\end{equation}
Thus, we obtain an effective non-Hermitian Hamiltonian evolution,
$H_{\mathrm{eff}}=M-\mathrm{i}\frac{\gamma}{2}$. The eigenvalues of
the above effective Hamiltonian under the weak coupling constant
approximation are:
\begin{equation}\label{eg1}
\omega_{+}=\omega_1-\mathrm{i}\pi|\lambda_1|^2+O(\lambda^4), ~~~
\omega_{-}=\omega_2-\mathrm{i}\pi|\lambda_2|^2+O(\lambda^4),
\end{equation}
In a first and very rough approximation, the eigenvectors of the
effective Hamiltonian are the same as the postulated kaons states.
\begin{equation}
| f_+\rangle=\left(\begin{array}{c} 1\\0
\end{array}\right)=|K_1\rangle ~~~\mathrm{and}~~~
| f_-\rangle=\left(\begin{array}{c} 0\\1
\end{array}\right)=|K_2\rangle,
\end{equation}
Phenomenology imposes that the complex Friedrichs energies $
\omega_{\pm}$ coincide with the observed complex energies. The
Friedrichs energies depend on the choice of the four parameters
$\omega_{1}$, $\omega_{2}$, $\lambda_{1}$ and $\lambda_{2}$ and the
observed complex energies are directly derived from the experimental
determination of four other parameters, the masses $m_{S}$ and
$m_{L}$ and the lifetimes $\tau_{S}$ and $\tau_{L}$. We must thus
adjust the theoretical parameters in order that they fit the
experimental data. This can be done by comparing the eigenvalue of
the effective matrix with the eigenvalue of the mass-decay matrix
which is taken in the expression (\ref{kkeigen}). Finally, we have
\begin{eqnarray} \label{fp9}
\nonumber& &\omega_1=m_S ,~~~2\pi|\lambda_1|^2=\Gamma_S,\\
& & \omega_2=m_L, ~~~2\pi|\lambda_2|^2=\Gamma_L.
\end{eqnarray}
The above identities yield
\begin{equation}\label{lambada}
\lambda_1=\sqrt{\frac{\Gamma_S}{2\pi}}\, e^{\mathrm{i}\theta_S},
~~~\lambda_2=\sqrt{\frac{\Gamma_L}{2\pi}}\, e^{\mathrm{i}\theta_L}
\end{equation}
where  $\theta_S$ and $\theta_L$ are real constants.

\emph{$CPT$ invariance:}
 Let us now discuss the $CPT$ invariance in our model. As
mentioned in the texts books like \cite{hokim,leebook}, $CPT$
invariance imposes some conditions on the mass-decay matrix, i.e.
\begin{equation}\label{cpt}
M_{11}=M_{22},~ \Gamma_{11}=\Gamma_{22},~ M_{12}=M^*_{21}~~\mathrm{
and}~~  \Gamma_{12}=\Gamma^*_{21}
\end{equation}
in the $\mathrm{K}^0$ and $\overline{\mathrm{K}}^0$ bases. But, we
note that our effective Hamiltonian is written in the $\mathrm{K}_1$
and $\mathrm{K}_2$ bases. Thus, we have to rewrite in the
$\mathrm{K}^0$ and $\overline{\mathrm{K}}^0$ bases. Thus, the
transformation matrix $T$ from the  $\mathrm{K}_1$ and
$\mathrm{K}_2$ bases to the $\mathrm{K}^0$ and
$\overline{\mathrm{K}}^0$ bases is obtained  as
\begin{equation}
T=\frac{1}{\sqrt 2}\left(\begin{array}{cc} 1&1\\1&-1
\end{array}\right)=T^{-1}.
\end{equation}
Then, the effective Hamiltonian in the $\mathrm{K}^0$ and
$\overline{\mathrm{K}}^0$  bases, $H_{\mathrm{eff}}^{0\overline{0}}$
is obtained by
\begin{equation}
H_{\mathrm{eff}}^{0\overline{0}}=TH_{\mathrm{eff}}T^{-1}=\frac{1}{2}\left(\begin{array}{cc}
1&1\\1&-1
\end{array}\right)
\left(\begin{array}{cc} \omega_1 - \mathrm{i} \pi |\lambda_1|^2 & -
\mathrm{i} \pi
\lambda_1^* \lambda_2\\
- \mathrm{i} \pi  \lambda_1\lambda_2^* & \omega_2 - \mathrm{i} \pi
|\lambda_2|^2
\end{array}\right)
\left(\begin{array}{cc} 1&1\\1&-1
\end{array}\right).\label{fh1-1}
\end{equation}
 we have, $H_{\mathrm{eff}}^{0\overline{0}} =$
\begin{equation}
\left(\begin{array}{cc} (m_S+ m_L) - \frac{\mathrm{i}}{2} \left(
\Gamma_S+\Gamma_L+2\sqrt{\Gamma_S\Gamma_L}\cos\triangle\theta\right),
&(m_S- m_L) - \frac{\mathrm{i}}{2} \left(
\Gamma_S-\Gamma_L+2\mathrm{i}\sqrt{\Gamma_S\Gamma_L}\sin\triangle\theta\right) \\
(m_S- m_L) - \frac{\mathrm{i}}{2} \left(
\Gamma_S-\Gamma_L-2\mathrm{i}\sqrt{\Gamma_S\Gamma_L}\sin\triangle\theta\right),
& (m_S+ m_L) - \frac{\mathrm{i}}{2} \left(
\Gamma_S+\Gamma_L-2\sqrt{\Gamma_S\Gamma_L}\cos\triangle\theta\right)
\end{array}\right)
.\label{fh1-2}
\end{equation}
where $\triangle\theta=\theta_L-\theta_S$. $CPT$ invariance
conditions in (\ref{cpt}) impose that
\begin{equation}
\triangle\theta=k\pi+\frac{\pi}{2},~~~ (k=\cdots,-1,0,1,\cdots).
\end{equation}
Here we choose $k=-1$, consequently,
$\triangle\theta=-\frac{\pi}{2}$. Then, we have
\begin{equation}
\begin{array}{ll}
M_{11}=M_{22}=(m_S+
m_L),&\Gamma_{11}=\Gamma_{22}=\Gamma_S+\Gamma_L,\\
M_{12}=M_{21}^*=(m_S-
m_L),&\Gamma_{12}=\Gamma_{21}^*=\Gamma_S-\Gamma_L-2\mathrm{i}\,\sqrt{\Gamma_S\Gamma_L}.
\end{array}
\end{equation}

\emph{$CP$-violation:} \label{section} Let us study in this case the
$CP$-violation. The Friedrichs model allows us to estimate the value
of $\epsilon$. For this purpose,  the effective Hamiltonian
(\ref{fh1}) acts on the $|\mathrm{K}_S\rangle$ vector states
(\ref{kk2}) as an eigenstate corresponding to the eigenvalue
$\omega_+=\omega_1-\mathrm{i}\pi|\lambda_1|^2=m_S-\mathrm{i}
\frac{\Gamma_S}{2}$, so that  we must impose that
$H_{\mathrm{eff}}|(f_++\epsilon
f_-)\rangle=H_{\mathrm{eff}}\big{(}^{1}_{\epsilon}\big{)}=\omega_+
\big{ (}^{1}_{\epsilon}\big{)}$, from which we obtain after
straightforward calculations that
\begin{equation}
\epsilon=\frac{\frac{1}{2}\sqrt{\Gamma_L\Gamma_S}}
{(m_L-m_S)-\frac{\mathrm{i}}{2}(\Gamma_L-\Gamma_S)}.\label{angle1}
\end{equation}
By using the experimental ratio
$\frac{(m_L-m_S)}{-(\Gamma_L-\Gamma_S)}\approx \triangle m
\tau_S\approx 0.47$ and the above experimental values of
$\Gamma_L,\Gamma_S, m_L$, $m_S$, we obtain the following estimated
value for $\epsilon$:
\begin{equation}\label{angle2}
\epsilon=\sqrt{\frac{\Gamma_L}{\Gamma_S}}~~
e^{\mathrm{i}(46.77)^\circ}=\sqrt{\frac{1.82\times 10^{-3}}{2}}~
e^{\mathrm{i}(46.77)^\circ}=14\,\epsilon^\mathrm{exp}.
\end{equation}
We see that the $\epsilon$ argument is the  same as the experimental
value but the magnitude of the $CP$-violation parameter is quite
larger than its experimental value.

The reason is that, as we have shown in a previous work \cite{cds3},
we did not normalize correctly the amplitudes associated to the two
interfering decay processes (short and long). In that work we solved
the problem by developing an  analogy between the temporal density
of decay and the spatial density of presence (this constitutes the
so-called wave function approach).

  \bigskip

Now we shall derive intensity formula for the meson decay
\cite{cds3} using the formalism of the time operator ($T'$) sketched
in  Section \ref{section2}.  By considering the relations
(\ref{approxpole}) and (\ref{etap2}) and supposing the $v(\omega)$
is a real function, we can write the $f_1(\omega)$ and
$f_2(\omega)$, the equations (\ref{af4}) and (\ref{af4-1}), as:
\begin{equation}\label{ww1}
f_i(\omega)=\frac{\sqrt{\gamma_i\over
2}e^{-\mathrm{i}\theta_i}}{\omega-\widetilde{\omega}_i+\frac{\mathbf{i}}{2}\gamma_i},
~~~(i=1,2),
\end{equation}
 where $\theta_i$ is the phase of the possibly complex coefficients $\lambda_i$.
By using the  Fourier transforms, i.e. equation (\ref{dc3}), for the
above equation, (\ref{ww1}), we obtain for $(i=1,2)$
\begin{equation}\label{ww2}
\hat{f}_i(\tau)=\left\{\begin{array}{ll} N \sqrt{\pi\,\gamma_i}\,
\,e^{-(\mathrm{i}\widetilde{\omega}_i+\frac{\gamma_i}{2})\tau-\mathrm{i}\theta_i},&
\tau\ge0\\0,&\tau<0
 \end{array}\right.
\end{equation}
where $N$ is the normalization constant. For $s=-\tau<0$,  we have
\begin{equation}\label{ww3}
\hat{f}_i(s)=\left\{\begin{array}{ll}
\sqrt{\pi\gamma_i}\,e^{(\mathrm{i}\widetilde{\omega}_i+\frac{\gamma_i}{2})s-\mathrm{i}\theta_i},&s\le0\\
\\0,&s>0.
 \end{array}\right.
\end{equation}
Finally, the normalization relation, i.e.
\begin{equation}\label{ww4}
\int_{-\infty}^{+\infty} ds\,|\hat{f}_i(s)|^2=1,~~~(i=1,2)
\end{equation}
yields:
\begin{equation}\label{ww5}
\hat{f}_i(s)=\left\{\begin{array}{ll}
\sqrt{\gamma_i}\,e^{(\mathrm{i}\widetilde{\omega}_i+\frac{\gamma_i}{2})s-\mathrm{i}\theta_i},&s\le0\\
\\0,&s>0.
 \end{array}\right.
\end{equation}
Here $f_i(s),~(i=1,2)$ is the form of the density of the
probability. Thus, the intensity is obtained by
\begin{eqnarray}\label{ww6}
\nonumber
I(s)&=&|C|^2|\epsilon_1f_1(s)+\epsilon_2f_2(s)|^2\\\nonumber
&=&I_0\left(e^{\gamma_1s}+|\epsilon|^2\,
\frac{\gamma_2}{\gamma_1}\,e^{\gamma_2s}
+|\epsilon|\sqrt{\frac{\gamma_2}{\gamma_1}}\,
e^{\frac{(\gamma_1+\gamma_2)}{2}s}
\cos((\widetilde{\omega}_1-\widetilde{\omega}_2)s+\theta_2-\theta_1+\arg(\epsilon))\right)\\
\end{eqnarray}
where $\epsilon=\epsilon_2/\epsilon_1$ and $C$ and
$I_0=|C|^2\epsilon_1^2\gamma_1$ are the constants.  This corresponds
to an effective value for $\epsilon$ that is no longer 14 times too
large as  in expression (\ref{angle2}) because it must be
renormalized. Identifying equations (\ref{intens}) and (\ref{ww6})
it is easy to show, as we have also done in \cite{cds3}, that,
$\epsilon^\mathrm{th}$, the theoretical prediction for the
experimental $CP$-violation parameter, obeys
\begin{equation}\label{re1}
\epsilon^\mathrm{th}=\epsilon\,\sqrt{\frac{\Gamma_L}{\Gamma_S}}=\frac{\Gamma_L}{\Gamma_S}\frac{\frac{1}{2}}
{\frac{\triangle
m}{\Gamma_S}-\mathrm{i}\,\frac{\triangle\gamma}{2\Gamma_S}}.
\end{equation}

Substituting in the expression (\ref{re1}) the physically observed
masses and lifetimes of the short and long kaon states we find that
$\epsilon^\mathrm{th}\approx 0.6 \, \epsilon^\mathrm{exp}$ which
constitutes an improvement in comparison to the non-renormalized
estimation (\ref{angle2}). We shall also reconsider similar results
in the case of $B$ and $D$ particles in a next section.

In the next coming section, we shall use the  time super-operator
($T$) formalism as a non Wigner-Weisskopf approximation method to
obtain the $CP$-violation parameter. This formalism also predicts a
$CP$-violation parameter comparable to the experimental value.


\section{Computation of spectral projections of \emph{T} in a
Friedrichs model}\label{sec4}

 In this section, we will compute the survival probability and
we obtain the theoretical   $CP$-violation parameters for the mesons
K, B and D. Then, we compare our results to the experimental
$CP$-violation parameters. We shall see that our theoretical results
provide a good estimation of the experimentally measured quantities.
Moreover, a fine structure appears in the case of kaons, which
brought us to conceive an experimental test aimed at falsifying the
time super-operator approach, that we shall discuss in the
conclusion.

 By considering $v(\omega)$ a real test function and using the
equation (\ref{etap2}) we obtain $\mathfrak{F}_{ji}(\nu,E)$ in the
following form:
\begin{equation}
\mathfrak{F}_{ji}(\nu,E)=\left\{ \begin{array}{ll}
 \frac{\lambda_j\lambda_i^*}{\nu^*_j(\nu+\nu_i)}& \mbox{ $\nu > 0$}\\\\
\frac{\lambda_j^*\lambda_i}{\nu_i(\nu^*_j-\nu)}& \mbox{ $\nu <0$}.
\end{array} \right.\label{c1}
 \end{equation}
where $i,j=1,2$ and
\begin{equation}\label{gss}
\nu_{i}:=a_{i}+\mathrm{i} b_{i} := (E-\widetilde{\omega}_{i})+
\mathrm{i} \frac{\gamma_{i}}{2}.
\end{equation}
For obtaining ${\cal P}_s\mathfrak{F}_{ij}(\nu,E) ~(s<0)$, we shall
use the formula (\ref{t1}). First we compute:
\begin{equation}\label{gs}
G_{ji}(\nu,E)=\mathbf{H}(e^{\mathrm{i}s
\nu}\mathfrak{F}_{ji})(\nu,E)= \frac{1}{\pi}
\textsf{P}\int_{-\infty}^{\infty} \frac{e^{\mathrm{i}s
x}\mathfrak{F}_{ji}(x,E)}{x-\nu} dx
\end{equation}
Now, we substitute (\ref{c1}) in (\ref{gs}), so we have
\begin{equation}
 G_{ji}(\nu,E)=
\frac{1}{\pi}\textsf{P}\left[\lambda_i\lambda_j^*\,\int_{-\infty}^0
\frac{e^{\mathrm{i}s
x}}{\nu_i(x-\nu)(\nu_j^*-x)}\,dx+\lambda_i^*\lambda_j\,\int_0^{+\infty}
\frac{e^{\mathrm{i}s x}}{\nu_j^*(x-\nu)(\nu_i+x)}\,dx\right]
\end{equation}
which for the $\nu>0$ has the following form:
\begin{equation}
G_{ji}(\nu,E)=\frac{1}{\pi}\left[\lambda_i\lambda_j^*\,\int_{-\infty}^0
\frac{e^{\mathrm{i}s
x}}{\nu_i(x-\nu)(\nu_j^*-x)}\,dx+\lambda_i^*\lambda_j\textsf{P}\int_0^{+\infty}
\frac{e^{\mathrm{i}s
x}}{\nu_j^*(x-\nu)(\nu_i+x)}\,dx\right].\label{t2}
\end{equation}
A complete computation  of the $G_{ii}(\nu,E)$ is showed in
\cite{cs4}. Finally, ${\cal P}_s\mathfrak{F}_{ij}(\nu,E)$ is
obtained as: for $i=j$
\begin{eqnarray}
\nonumber\mathcal{P}_s\mathfrak{F}_{ii}(\nu,E) =
\mathrm{i}|\lambda_i|^2\,e^{-\mathrm{i}s\nu}\bigg{[}\frac{-1}
{2\pi\nu_i(\nu_i^*-\nu)}\bigg{(}\int_{-\infty}^0
\frac{e^{-sy}}{y+\mathrm{i}\nu_i^*}dy- \int_{-\infty}^0
\frac{e^{-sy}}{y+\mathrm{i}\nu}dy\bigg{)}& &  \\
\nonumber +\frac{1} {2\pi\nu_i^*(\nu+\nu_i)}\bigg{(}\int_{-\infty}^0
\frac{e^{-sy}}{y-\mathrm{i}\nu_i}dy- \int_{-\infty}^0
\frac{e^{-sy}}{y+\mathrm{i}\nu}dy\bigg{)}\bigg{]}& & \\
+\left\{
\begin{array}{ll}
 |\lambda_i|^2\,e^{-\mathrm{i}s\nu }[\frac{e^{\mathrm{i}s\nu_i^* }
}{\nu_i(\nu_i^*-\nu)}-\frac{e^{-\mathrm{i}s\nu_i }
}{\nu_i^*(\nu_i+\nu)}],& E< \widetilde{\omega}_1 \\  \\ 0, &
E>\widetilde{\omega}_1.
\end{array} \right.\label{t5}
\end{eqnarray}
and  by considering $\widetilde{\omega}_i<\widetilde{\omega}_j,\,\,
\mathfrak{F}_{ij}$ for $i\neq j$ have the following form :
\begin{eqnarray}
\nonumber\mathcal{P}_s\mathfrak{F}_{ji}(\nu,E) =
\mathrm{i}\,e^{-\mathrm{i}s\nu}\bigg{[}\frac{-\lambda_i\lambda_j^*}
{2\pi\nu_i(\nu_j^*-\nu)}\bigg{(}\int_{-\infty}^0
\frac{e^{-sy}}{y+\mathrm{i}\nu_j^*}dy- \int_{-\infty}^0
\frac{e^{-sy}}{y+\mathrm{i}\nu}dy\bigg{)}& &  \\
\nonumber +\frac{\lambda_i^*\lambda_j}
{2\pi\nu_j^*(\nu+\nu_i)}\bigg{(}\int_{-\infty}^0
\frac{e^{-sy}}{y-\mathrm{i}\nu_i}dy- \int_{-\infty}^0
\frac{e^{-sy}}{y+\mathrm{i}\nu}dy\bigg{)}\bigg{]}& & \\
+\left\{
\begin{array}{ll}
e^{-\mathrm{i}s\nu }[\frac{\lambda_i\lambda_j^*e^{\mathrm{i}s\nu_j^*
}
}{\nu_i(\nu_j^*-\nu)}-\frac{\lambda_i^*\lambda_je^{-\mathrm{i}s\nu_i
}
}{\nu_j^*(\nu_i+\nu)}],& E< \widetilde{\omega}_i \\ \\
\lambda_i\lambda_j^*\,e^{-\mathrm{i}s\nu
}\,\frac{e^{\mathrm{i}s\nu_i^* } }{\nu_i(\nu_j^*-\nu)},&
\widetilde{\omega}_i<E< \widetilde{\omega}_j \\
\\
0, & E>\widetilde{\omega}_j.
\end{array} \right.\label{t5-2}
\end{eqnarray}
 In the equations (\ref{t5}) and (\ref{t5-2}) the non-integral terms yield the poles and lead
to the resonance, and the integral terms yield an algebraical term
analog to the background in the Hamiltonian theories \cite{bohm1}.
We can also compute the same result for the case $\nu<0$. We will
neglect the the background (the integrals terms). Then, the above
equation is rewritten as:
\begin{eqnarray}
\nonumber\mathcal{P}_s\mathfrak{F}_{ii}(\nu,E) \simeq \left\{
\begin{array}{ll}
 |\lambda_i|^2\,e^{-\mathrm{i}s\nu }[\frac{e^{\mathrm{i}s\nu_i^* }
}{\nu_i(\nu_i^*-\nu)}-\frac{e^{-\mathrm{i}s\nu_i }
}{\nu_i^*(\nu_i+\nu)}],& E\leq \widetilde{\omega}_1 \\ \\
0, & E>\widetilde{\omega}_1.
\end{array} \right.\label{t53}
\end{eqnarray}
and for $i\neq j$
\begin{eqnarray}
\nonumber\mathcal{P}_s\mathfrak{F}_{ij}(\nu,E)\simeq \left\{
\begin{array}{ll}
e^{-\mathrm{i}s\nu }[\frac{\lambda_i\lambda_j^*e^{\mathrm{i}s\nu_j^*
}
}{\nu_i(\nu_j^*-\nu)}-\frac{\lambda_i^*\lambda_je^{-\mathrm{i}s\nu_i
}
}{\nu_j^*(\nu_i+\nu)}],& E\leq \widetilde{\omega}_i \\ \\
\lambda_i\lambda_j^*\,e^{-\mathrm{i}s\nu
}\,\frac{e^{\mathrm{i}s\nu_i^* }
}{\nu_i(\nu_j^*-\nu)},& \widetilde{\omega}_i<E\leq \widetilde{\omega}_j \\ \\
0, & E>\widetilde{\omega}_j.
\end{array} \right.\label{t54}
\end{eqnarray}

Now, we would compute the survival  probability, i.e.
\begin{equation}
p_\rho(s)=\|\mathfrak{P}_\rho(s)\|=\|\,|\epsilon_1|^2\mathcal{P}_s\mathfrak{F}_{11}(\nu,E)
+\epsilon_1\epsilon^*_2\mathcal{P}_s\mathfrak{F}_{12}(\nu,E)
+\epsilon_2\epsilon^*_1\mathcal{P}_s\mathfrak{F}_{21}(\nu,E)
+|\epsilon_2|^2\mathcal{P}_s\mathfrak{F}_{22}(\nu,E)\|
\end{equation}
where $\|\cdot\|=\int_{0}^\infty dE\,\int_{-\infty}^\infty d\nu\,
|\cdot|^2$. We  see that $\mathfrak{P}_\rho(s)$  can be written as :
\begin{equation}\label{p1}
\mathfrak{P}_\rho(s)\simeq\left\{
\begin{array}{ll}
 e^{-\mathrm{i}s\nu
}\bigg{[}\left(\frac{\epsilon_1^*\lambda_1}{\nu_1}+\frac{\epsilon_2^*\lambda_2}{\nu_2}
\right)\left(\frac{\epsilon_1\lambda_1^*\,e^{\mathrm{i}s \nu_1^*}
}{\nu_1^*-\nu}+\frac{\epsilon_2\lambda_2^*\,e^{\mathrm{i}s \nu_2^*}
}{\nu_2^*-\nu}\right)\\
\,\,\,\,\,\,\,\,\,\,\,\,\,\,\,-\left(\frac{\epsilon_1\lambda_1}{\nu_1^*}+\frac{\epsilon_2\lambda_2}{\nu_2^*}
\right)\left(\frac{\epsilon_1^*\lambda_1^*\,e^{-\mathrm{i}s \nu_1}
}{\nu_1+\nu}+\frac{\epsilon_2^*\lambda_2^*\,e^{-\mathrm{i}s \nu_2}
}{\nu_2+\nu}\right)\bigg{]} &E\leq \widetilde{\omega}_1,\\ \\
e^{-\mathrm{i}s\nu
}\bigg{[}\left(\frac{\epsilon_1^*\lambda_1}{\nu_1}+\frac{\epsilon_2^*\lambda_2}{\nu_2}
\right)\frac{\epsilon_2\lambda_2^*\,e^{\mathrm{i}s \nu_2^*}
}{(\nu_2^*-\nu)}
-\left(\frac{\epsilon_1\lambda_1}{\nu_1^*}+\frac{\epsilon_2\lambda_2}{\nu_2^*}
\right)\frac{\epsilon_2^*\lambda_2^*\,e^{-\mathrm{i}s \nu_2}
}{(\nu_2+\nu)}\bigg{]},&\widetilde{\omega}_1<E\leq
\widetilde{\omega}_2\\ \\
0,& E>\widetilde{\omega}_2\\
\end{array}\right.
\end{equation}

Now, by remembering that $b_i=|\lambda_i|^2,~(i=1,2)$,  the square
norm of $\mathfrak{P}_\rho(s)$ is obtained as:
\begin{equation}\label{p3}
|\mathfrak{P}_\rho(s)|^2\simeq\left\{
\begin{array}{ll}
\left|\frac{\epsilon_1\lambda_1}{\nu_1}+\frac{\epsilon_2\lambda_2}{\nu_2}
\right|^2\bigg{[}\frac{|\epsilon_1|^2|\lambda_1|^2\,e^{2b_1s}
}{|\nu_1^*-\nu|^2}+\frac{|\epsilon_2|^2|\lambda_2|^2\,e^{2b_2s}
}{|\nu_2^*-\nu|^2}+\frac{|\epsilon_1|^2|\lambda_1|^2\,e^{2b_1s}
}{|\nu_1+\nu|^2}+\frac{|\epsilon_2|^2|\lambda_2|^2\,e^{2b_2s} }{|\nu_2+\nu|^2}\\
\,\,\,\,\,+\,e^{(b_1+b_2)s}\left(\frac{\epsilon_1\epsilon_2^*\lambda_1^*\lambda_2e^{\mathrm{i}(a_1-a_2)s}
}{(\nu_1^*-\nu)(\nu_2-\nu)}+\frac{\epsilon_1\epsilon_2^*\lambda_1^*\lambda_2e^{\mathrm{i}(a_1-a_2)s}
}{(\nu_1^*+\nu)(\nu_2+\nu)}+\mathrm{C.C.}\right)\bigg{]},&E\leq
\widetilde{\omega}_1\\ \\
\left|\frac{\epsilon_1\lambda_1}{\nu_1}+\frac{\epsilon_2\lambda_2}{\nu_2}
\right|^2\bigg{[}\frac{|\epsilon_2|^2|\lambda_2|^2\,e^{2b_2s}
}{|\nu_2^*-\nu|^2}+\frac{|\epsilon_2|^2|\lambda_2|^2\,e^{2b_2s}
}{|\nu_2+\nu|^2}\bigg{]},&\widetilde{\omega}_1<E\leq
\widetilde{\omega}_2\\ \\ 0,& E> \widetilde{\omega}_2
\end{array}\right.
\end{equation}
where  the terms that oscillate with a frequency equal to the
difference of the two masses, i.e.
$(\widetilde{\omega}_2-\widetilde{\omega}_1)$ is kept, the other
decay terms oscillating with the frequency of  one of the masses
only are neglected since we have in the weak-coupling regime and the
high-mass.

The integral over $\nu$  arrives at:
\begin{equation}\label{p5}
\int_{-\infty}^{\infty} d\nu\,|\mathfrak{P}_\rho(s)|^2\simeq\left\{
\begin{array}{ll}
2\pi\left|\frac{\epsilon_1\lambda_1}{\nu_1}+\frac{\epsilon_2\lambda_2}{\nu_2}
\right|^2\bigg{[}|\epsilon_1|^2e^{2b_1s}
+|\epsilon_2|^2e^{2b_2s}\\~~~~~~~~+\left(
\frac{2\mathrm{i}\epsilon_1^*\epsilon_2\lambda_1^*\lambda_2\,e^{(b_1+b_2)s}\,e^{\mathrm{i}
( \widetilde{\omega}_1- \widetilde{\omega}_2)s} }{
(\widetilde{\omega}_2- \widetilde{\omega}_1)
+\mathrm{i}(b_1+b_2)}+\mathrm{C.C.}\right)\bigg{]},&E\leq
\widetilde{\omega}_1 \\ \\
2\pi\left|\frac{\epsilon_1\lambda_1}{\nu_1}+\frac{\epsilon_2\lambda_2}{\nu_2}
\right|^2\,|\epsilon_2|^2\,e^{2b_2s},&\widetilde{\omega}_1<E\leq\widetilde{\omega}_2\\ \\
0,&E>\widetilde{\omega}_2
\end{array}\right.
\end{equation}
Only the terms of the square norm are depended to $E$ and we have
\begin{equation}\label{p6}
\left|\frac{\epsilon_1\lambda_1}{\nu_1}+\frac{\epsilon_2\lambda_2}{\nu_2}
\right|^2=
\left|\frac{\epsilon_1\lambda_1}{E-\widetilde{\omega}_1+\mathrm{i}b_1}\right|^2+\left|
\frac{\epsilon_2\lambda_2}{E-\widetilde{\omega}_2+\mathrm{i}b_2}
\right|^2+\left(
\frac{\epsilon_1^*\lambda_1^*}{(E-\widetilde{\omega}_1+\mathrm{i}b_1)}\frac{\epsilon_2\lambda_2}{(E-\widetilde{\omega}_2-\mathrm{i}b_2)}
+\mathrm{C.C.}\right)
\end{equation}
The integral over $E$ of the above expression is like the following
integrals
\begin{equation}
\int dE \left|\frac{\sqrt{b_i}}{(E-\widetilde{\omega}_i)+
\mathrm{i}b_i}\right|^2 = \arctan
\left(\frac{E-\widetilde{\omega}_i}{b_i}\right)
\end{equation}
and
\begin{eqnarray}
\nonumber& & \int dE \frac{\lambda_1^* \lambda_2}{(x-a_1+
\mathrm{i}b_1)(E-a_2- \mathrm{i}b_2)} =\frac{-\lambda_1^*\lambda_2}{
(\widetilde{\omega}_2- \widetilde{\omega}_1)
+\mathrm{i}(b_1+b_2)}\bigg{(} \mathrm{i}\arctan
\frac{b_1}{E-\widetilde{\omega}_1}\\& & +\mathrm{i}\arctan
\frac{b_2}{E-\widetilde{\omega}_2}+\log\sqrt{(E-\widetilde{\omega}_1)^2+b_1^2}-
\log\sqrt{(E-\widetilde{\omega}_2)^2+b_2^2}\bigg{)}
\end{eqnarray}
Now, we integrate from equation(\ref{p6}) over $E$ from $0$ to
$\infty$. Firstly, for the interval $E\in[0,\widetilde{\omega}_1]$
we have
\begin{eqnarray}\label{p7}
\nonumber\mathfrak{I}_1&=&\int_0^{\widetilde{\omega}_1} dE\,
\left|\frac{\epsilon_1\lambda_1}{\nu_1}+\frac{\epsilon_2\lambda_2}{\nu_2}
\right|^2 = |\epsilon_1|^2\arctan\frac{\widetilde{\omega}_1}{b_1}
+|\epsilon_2|^2\left(\arctan\frac{\widetilde{\omega}_2-\widetilde{\omega}_1}{b_2}+
\arctan\frac{\widetilde{\omega}_1}{b_2}\right)
\\ \nonumber&-&\bigg{[}\left(\frac{\epsilon_1^*\epsilon_2\sqrt{b_1b_2}}{(\widetilde{\omega}_1-\widetilde{\omega}_2)
+\mathrm{i}(b_1+b_2)}\right)
\bigg{(}\mathrm{i}(\frac{\pi}{2}+\arctan\frac{b_1}{\widetilde{\omega}_1}
+\arctan\frac{b_2}{\widetilde{\omega}_2}\\&+&\arctan\frac{b_2}{\widetilde{\omega}_2-\widetilde{\omega}_1})
+\frac{1}{2}\log\frac{b_1^2(\widetilde{\omega}_2^2+b_2^2)}{(\widetilde{\omega}_1^2+b_1^2)
((\widetilde{\omega}_2-\widetilde{\omega}_1)^2
+b_2^2)}\bigg{)}+\mathrm{C.C.}\bigg{]}.
\end{eqnarray}
For $E\in]\widetilde{\omega}_1,\widetilde{\omega}_2]$ we have
\begin{eqnarray}\label{p8}
\nonumber\mathfrak{I}_2&=&\int_{\widetilde{\omega}_1}^{\widetilde{\omega}_2}
dE\,
\left|\frac{\epsilon_1\lambda_1}{\nu_1}+\frac{\epsilon_2\lambda_2}{\nu_2}
\right|^2 =
|\epsilon_1|^2\arctan\frac{\widetilde{\omega}_2-\widetilde{\omega}_1}{b_1}
+|\epsilon_2|^2\arctan\frac{\widetilde{\omega}_2-\widetilde{\omega}_1}{b_2}
\\ \nonumber&-&\bigg{[}\left(\frac{\epsilon_1^*\epsilon_2\sqrt{b_1b_2}}{(\widetilde{\omega}_1-\widetilde{\omega}_2)
+\mathrm{i}(b_1+b_2)}\right)
\bigg{(}\mathrm{i}(\arctan\frac{b_1}{\widetilde{\omega}_2-\widetilde{\omega}_1}-
\arctan\frac{b_2}{\widetilde{\omega}_2-\widetilde{\omega}_1})\\&+&
\frac{1}{2}\log\frac{b_1^2b_2^2}{((\widetilde{\omega}_2-\widetilde{\omega}_1)^2+b_1^2)
((\widetilde{\omega}_2-\widetilde{\omega}_1)^2
+b_2^2)}\bigg{)}+\mathrm{C.C.}\bigg{]}.
\end{eqnarray}


\subsection{K-meson}

For the weak-coupling constants we have
$b_i\ll\widetilde{\omega}_i,~(i=1,2)$ and also by supposing
$\widetilde{\omega}_1\sim\widetilde{\omega}_2$,
$(\widetilde{\omega}_2-\widetilde{\omega}_1)\sim b_1$  and
$\frac{b_2}{b_1}\ll1$, we have
\begin{eqnarray}\label{p9}
\mathfrak{I}_1&\simeq&\frac{\pi}{2}\left(|\epsilon_1|^2+2|\epsilon_2|^2+
\left(\frac{\epsilon_1^*\epsilon_2\lambda_1^*\lambda_2}{(\widetilde{\omega}_1-\widetilde{\omega}_2)
+\mathrm{i}(b_1+b_2)}+\mathrm{C.C.}\right)\right)\approx\frac{\pi}{2}\\
\mathfrak{I}_2&\simeq&\frac{\pi
}{4}\left(|\epsilon_1|^2+2|\epsilon_2|^2+\left(\frac{\epsilon_1^*\epsilon_2\lambda_1^*\lambda_2}{(\widetilde{\omega}_1-\widetilde{\omega}_2)
+\mathrm{i}(b_1+b_2)}+\mathrm{C.C.}\right)\right)\approx\frac{\pi}{4}
\end{eqnarray}
where we used the normalization relation, i.e.
$(|\epsilon_1|^2+|\epsilon_2|^2)=1$.

Finally, we obtain
\begin{eqnarray}\label{10}
\nonumber p_\rho(s)&\simeq
&\frac{\pi}{2}\bigg{[}|\epsilon_1|^2e^{2b_1s}
+\frac{3}{2}|\epsilon_2|^2\,e^{2b_2s}+\left(
\frac{\mathrm{i}\epsilon_1^*\epsilon_2\lambda_1^*\lambda_2\,e^{(b_1+b_2)s}\,e^{\mathrm{i}
( \widetilde{\omega}_1- \widetilde{\omega}_2)s}}{
(\widetilde{\omega}_2- \widetilde{\omega}_1)
+\mathrm{i}(b_1+b_2)}+\mathrm{C.C.}\right)\bigg{]}\\ \nonumber
&\simeq&\frac{\pi}{2}|\epsilon_1|^2\bigg{[}e^{2b_1s}
+\frac{3}{2}|\epsilon|^2\,e^{2b_2s}+\left(
\frac{\mathrm{i}\epsilon\lambda_1^*\lambda_2\,e^{(b_1+b_2)s}\,e^{\mathrm{i}
( \widetilde{\omega}_1- \widetilde{\omega}_2)s} }{
(\widetilde{\omega}_2- \widetilde{\omega}_1)
+\mathrm{i}(b_1+b_2)}+\mathrm{C.C.}\right)\bigg{]}.\\
\end{eqnarray}
where
\begin{equation}\label{s3}
\epsilon
=|\epsilon|\,e^{\mathrm{i}\phi}:=\frac{\epsilon_2}{\epsilon_1}.
\end{equation}
The derivative of the equation (\ref{10})  yields the time
super-operator density of the probability or  intensity:
 \begin{equation}\label{da1}
 I(s):=\frac{d p_\rho(s)}{ds}= I_0\bigg{[}e^{2b_1s}
+|\epsilon|^2\frac{3}{2}\frac{b_2}{b_1}\,e^{2b_2s}+2
|\epsilon|\,\sqrt{\frac{b_2}{b_1}}\,e^{(b_1+b_2)s}\,\cos ((
\widetilde{\omega}_1-
\widetilde{\omega}_2)s+\phi+\theta_2-\theta_1)\bigg{]}.
\end{equation}
where $I_0=(\pi|\epsilon_1|^2b_1)/2$ and
$\lambda_i=\sqrt{b_i}\,e^{\mathrm{i}\theta_i},~(i=1,2)$. This
expression differs by $\frac{3}{2}$ term from the intensity derived
previous by \cite{cds3}  from the integrated probability of decay of
two exponentially decay process or relation (\ref{re1}) that we
obtained  in the Hilbert space which we call the time operator
prediction.

Let us now evaluate the predictions related to the above equation in
the different time intervals and let us compare them with the
intensity introduced in the equation (\ref{intens}). Firstly, for
$t=-s\sim10\times\tau_S$ or $t\gg\tau_S$ which that the term
effective is: $|\epsilon|^2\frac{3}{2}\frac{b_2}{b_1}\,e^{2b_2s}$
and comparing for the same time  with the equation (\ref{intens})
yields the $CP$-violation parameter is
$|\epsilon|^2\frac{3}{2}\frac{b_2}{b_1}$. Thus, the equations
(\ref{da1}) and (\ref{intens}) for $t=-s\gg\tau_S$ can be written
approximately as
\begin{eqnarray}\label{10-1}
I(s)\approx
I_0\left|\epsilon^\mathrm{th}\right|^2\,e^{2b_2s}~~~\mathrm{and}~~~
I(t)\approx
I_0\left|\epsilon^\mathrm{exp}\right|^2\,e^{-\gamma_Lt},~~~(-s=t\gg\tau_S)
\end{eqnarray}
where
\begin{equation}\label{epsth}
\epsilon^\mathrm{th}=\epsilon\,\sqrt{\frac{3}{2}\frac{b_2}{b_1}}
\end{equation}
and the coefficient $\sqrt{\frac{3b_2}{2b_1}}$ in the above equation
is the correction which is  obtained by the time operator formalism
and by using the condition
$(\widetilde{\omega}_2-\widetilde{\omega}_1)\sim b_1\neq0$, then
$\mathfrak{I}_2\neq0$. Secondly, for the time of the order oft
$\tau_S$ ($t<5\tau_S$) we have
\begin{equation}\label{ee}
    I(s)\approx I_0\,e^{2b_1s}~~~\mathrm{and} ~~~
    I(t)\approx
    I_0\,e^{-\gamma_St},~~~(t<5\tau_S)
\end{equation}
Finally, for intermediate times ($5\tau_S< t<10\tau_S$) we have
\begin{eqnarray}\label{ee1}
 \nonumber  I(s)&\approx& I_0|\epsilon|^2\frac{b_2}{b_1}\,e^{2b_2s}\,\cos ((
\widetilde{\omega}_1- \widetilde{\omega}_2)s+\phi+\theta_2-\theta_1)
~~~\mathrm{and}\\ I(t)&\approx&
I_0\left|\epsilon^\mathrm{exp}\right|^2\,
   e^{-\left(\frac{\Gamma_S+\Gamma_L}{2}\right)t}\,\cos ((
m_L- m_L)s+\arg(\epsilon^{\mathrm{exp}}))
\end{eqnarray}
The equations (\ref{10-1}), (\ref{ee}), (\ref{ee1}) and
(\ref{lambada})  yield
\begin{equation}\label{s1}
\begin{array}{ll}
b_1=\frac{\gamma_1}{2}=\frac{\Gamma_S}{2}=\frac{1}{2\tau_S}, & \widetilde{\omega}_1=m_S, \\
 b_2=\frac{\gamma_2}{2}=\frac{\Gamma_L}{2}
=\frac{1}{2\tau_L}, & \widetilde{\omega}_2=m_L, \\
\theta_1=\theta_S,&\theta_2=\theta_L
 \end{array}
\end{equation}
 The
$\epsilon$ is obtained in (\ref{angle2}), thus, we have
\begin{equation}\label{epp}
\epsilon^\mathrm{th}=\left(\epsilon\,\sqrt{\frac{3}{2}\frac{\Gamma_L}{\Gamma_S}}\,e^{-\mathrm{i}\frac{\pi}{2}}\right)=\sqrt{\frac{3}{2}}
\frac{\Gamma_L}{\Gamma_S}\frac{\frac{1}{2}} {\frac{\triangle
m}{\Gamma_S}-\mathrm{i}\,\frac{\triangle\gamma}{2\Gamma_S}}
\end{equation}
where $\triangle m=(m_L-m_S)$ and
$\triangle\gamma=(\Gamma_L-\Gamma_S)$. Then, by replacing the
experimental data we have
\begin{equation}\label{eppp}
\epsilon^\mathrm{th}=1.62\times10^{-3} \,
e^{\mathrm{i}(46.77^\circ)}=0.73\, \epsilon^{\mathrm{exp}}
\end{equation}

\subsection{B-meson}

It easy to see that the integral $\mathfrak{I}_2$, for the B-mesons
and D-mesons, is zero. So the intensity is written as:
\begin{eqnarray}\label{10-2}
 I(s)= I_0\bigg{[}e^{2b_1s}
+|\epsilon_1^\mathrm{th}|^2\,e^{2b_2s}+2
|\epsilon_1^\mathrm{th}|\,e^{(b_1+b_2)s}\,\cos ((
\widetilde{\omega}_1- \widetilde{\omega}_2)s+\phi)\bigg{]}
\end{eqnarray}
where
\begin{equation}\label{epsth1}
\epsilon^\mathrm{th}=\epsilon\,\sqrt{\frac{b_2}{b_1}}\,
e^{-\mathrm{i}\frac{\pi}{2}}
\end{equation}
This expression is the same, in the case of $B$ and $D$ particles in
the time operator and in the super-operator approaches, and the
theoretically estimated $CP$-violation parameter obeys the following
equation
\begin{equation}\label{ep}
\epsilon^\mathrm{th}=\epsilon\,\sqrt{\frac{\Gamma_L}{\Gamma_S}}\,e^{-\mathrm{i}\frac{\pi}{2}}=
\frac{\Gamma_L}{\Gamma_S}\frac{\frac{1}{2}} {\frac{\triangle
m}{\Gamma_S}-\mathrm{i}\,\frac{\triangle\gamma}{2\Gamma_S}}
\end{equation}
 which is not true in the case of $K$ particles. Also for
$B$ and $D$ particles the agreement with observations is quite good
as we shall now check.

Another example  is the  $CP$-violation in the decay of
$\mathrm{B}^0_s$  and $\overline{\mathrm{B}}^0_s$. The experimental
values are \cite{amslerB}
\begin{equation}\label{bb1}
\frac{\triangle\Gamma_s}{2\Gamma_s}=0.069^{+0.058}_{-0.062},~~~~\frac{1}{\Gamma_s}=1.470^{+0.026}_{-0.027}\,\,\mathrm{ps},
\end{equation}
or equivalently ($\Gamma_{L,H}=\Gamma_s\pm\triangle\Gamma_s/2$),
\begin{equation}\label{bb2}
\frac{1}{\Gamma_L}=1.419^{+0.039}_{-0.038}\,\,\mathrm{ps},~~~~\frac{1}{\Gamma_H}=1.525^{+0.062}_{-0.063}\,\,\mathrm{ps},
\end{equation}
and the difference of masses is
\begin{equation}\label{bb3}
\triangle m=17.7^{+6.4}_{-2.1}\,\,\mathrm{ps}^{-1}
\end{equation}
and the experimental $CP$-violation parameter of the $\mathrm{B}$
meson is \cite{amslerB,amslerCP}:
\begin{equation}\label{bb4}
\mathcal{A}_{SL}^{\mathrm{exp}}\simeq4\mathcal{R}e(\epsilon^{\mathrm{exp}}_B)=(-0.4\pm5.6)\times10^{-3}\Rightarrow
\left|\frac{q}{p}\right|^{\mathrm{exp}}=1.0002\pm0.0028.
\end{equation}
where
$\frac{\mathcal{A}_{SL}^{\mathrm{exp}}}{2}\approx1-\left|\frac{q}{p}\right|^{\mathrm{exp}}$.
By replacing in the equation (\ref{ep}) we obtain:
\begin{equation}\label{bb5}
\epsilon^{\mathrm{th}}_B=\frac{\Gamma_{L}}{\Gamma_{H}}\,
\frac{\frac{1}{2}}{\frac{\triangle
m}{\Gamma_s}-\mathrm{i}\frac{\triangle\Gamma_s}{2\Gamma_s}}=0.018 +
0.047\times10^{-3} \, \mathrm{i}
\end{equation}
Thus, our theoretical $\left|\frac{q}{p}\right|^{\mathrm{th}}$
prediction is:
\begin{equation}\label{bb6}
\left|\frac{q}{p}\right|^{\mathrm{th}}=\left|\frac{1-\epsilon^{\mathrm{th}}}{1+\epsilon^{\mathrm{th}}}\right|=0.96
\end{equation}
which is in fairly good agreement with the experimental value.


\subsection{D-meson}
The other example is the $CP$-violation in the decay of $\mathrm{D}$
meson. The experimental values for $CP$-violation of
$\mathrm{D}^0\rightarrow\mathrm{K}_S^0\,\pi^+\,\pi^-$ as reported by
Belle \cite{amslerD} are as follows:
\begin{eqnarray}\label{dd1}
\frac{\triangle\Gamma}{2\Gamma}=\left(0.37\pm0.25^{+0.07+0.07}_{-0.13-0.08}\right),\\
\frac{\triangle
m}{\Gamma}=\left(0.81\pm0.30^{+0.10+0.09}_{-0.07-0.16}\right)
\end{eqnarray}
where  $1/\Gamma=\tau, ~(\hbar=1)$ is the mean life time
\begin{equation}\label{dd2}
\frac{1}{\Gamma}=\tau=\frac{\tau_{\overline{\mathrm{D}}^0}+\tau_{\mathrm{D}^0}}{2}=(410.1\pm1.5)\times
10^{-3}\,\mathrm{ps}
\end{equation}
The  $CP$-violation parameters are experimentally denoted by
$\left(\frac{q}{p}\right)$ and given by:
\begin{equation}\label{cp111}
\left|\frac{q}{p}\right|^{\mathrm{exp}}=\left|\frac{1-\epsilon^{\mathrm{exp}}}{1+\epsilon^{\mathrm{exp}}}\right|
=\left(0.86^{+0.30+0.06}_{-0.29-0.03}\right)
\end{equation}
 and
 \begin{equation}\label{cp112}
\phi^{\mathrm{exp}}=\arg\left(\frac{q}{p}\right)^{\mathrm{exp}}
=\arg\left(\frac{1-\epsilon^{\mathrm{exp}}}{1+\epsilon^{\mathrm{exp}}}\right)
=\left(-14^{+16+5+2}_{-18-3-4}\right)^\circ.
\end{equation}
 By replacing in the expression (\ref{ep}) we obtain
 \begin{equation}
\epsilon^{\mathrm{th}}=\left(0.077+ 0.035 \mathrm{i}\right).
 \end{equation}
Consequently,
 \begin{equation}
\left|\frac{q}{p}\right|^{\mathrm{th}}=0.86,~~~\phi^{\mathrm{th}}=-4.02^\circ.
\end{equation}
which is once again in fairly good agreement with the experimental
value.

\section{Concluding remarks}
\subsection*{About the relevance and novelty of our results}
As we can see, the accuracy of the prediction (\ref{epp}) is
comparable to the one that we derived within the Wigner-Weisskopf
approach (\ref{re1}) (time operator instead of time super-operator).
Now, as we said before, the present results were derived under the
assumption that the spectrum of the continuous mode was not bounded
by below (no cut-off). In a precedent publication \cite{cds2}, we
considered the  Friedrichs model with a Gaussian factor form and
energy bounded by below (the spectrum of the continuous mode was
assumed there to vary from $0$ to $+\infty$). We showed that by
introducing a cut-off in the coupling between discrete and
continuous modes, the estimated value of $\epsilon$ slightly
differs, depending on the shape that we impose to the cut-off.
Therefore  a fine tuning of the estimated $CP$-violation parameter
is possible provided that the factor form is chosen conveniently.
Considered so,  the precision of the agreement with the measured
value of the $CP$-violation parameter is not very  convincing by it
self  (3 times the experimental value of the kaon $CP$-violation
parameter \cite{cds2}). What is convincing in our approach is that
we obtain the right order of magnitude for the K, B and D particles
altogether.


\subsection*{A crucial experiment for testing the validity of the Time Super-Operator ($T$) Formalism}
The most important novelty of the time operator approach is, in our
eyes, that it predicts that the distribution in time of the measured
populations of pions pairs significantly differs from the
predictions that could be made in the standard approach and/or in
the Wigner-Weisskopf approach provided we make a fit over the full
distribution (which means not only for times larger than the
lifetime of the ``Short" state but also for times comparable to it).
Indeed, taking account of the three contributions of the
distribution, which are the purely exponential, "Short" and "Long"
contributions, and the oscillating contribution, one sees that the
expression (\ref{da1}) radically differs from the expressions
(\ref{intens}) and (\ref{ww6}). This is due to the presence of the
coefficient $\frac{3b_2}{2b_1}$ in the above equation which is the
correction  obtained by the time super-operator formalism and by
using the condition $(\widetilde{\omega}_2-\widetilde{\omega}_1)\sim
b_1,$ that is, $(m_L-m_s)\sim\Gamma_S$ for kaons. Since in the case
of the  B and D mesons, no such relation  exists, the formula
obtained here coincide with the one derived using the
Wigner-Weisskopf  time operator approach \cite{cds3}.

So, one can conceive crucial experiments that would allow to falsify
the time operator approach and do not radically differ from the
original Christenson experiment. These experiments require to
measure the population of pairs of pions over a large range of times
(distances to the source), and to check whether the best fit is
provided by the expression (\ref{epsth}) or by the expressions
(\ref{intens}) and (\ref{ww6}).

In principle these effects will be tested on the LHC at CERN in the
coming months (years) so that the crucial experiment that we propose
here is feasible in the future.

\bigskip
\subsection*{Concluding remark}
 The formalism of the mass-decay matrix for the kaon decay was
introduced by LOY \cite{loy}. Then several other authors
\cite{chmisrasud,khalfin57,cds1} improved this model. The LOY model
requires the Wigner-Weisskopf  approximation, i.e. it requires to
assume that the energy interval varies from $-\infty$ to $+\infty$
and also that the coupling between discrete and continuous modes is
not restricted by a factor form or cut-off.

In \cite{cds1}, we used the 2-level Friedriche model and the
Wigner-Weisskopf approach  to obtain a mass-decay matrix. This
approach was improved by using a new concept of probability decay
density for mesons in \cite{cds3}. Beyond the Wigner-Weisskopf
approximation, we used the Friedrichs model with a cutoff that
amounts to bound from below the energy spectrum of the Hamiltonian
\cite{cds2}. In the present paper, we derived the decay probability
density in the formalism of the time super-operator, that also goes
beyond the Wigner-Weisskopf approximation.


\end{document}